\documentclass[aps,prb,a4paper,twocolumn]{revtex4}
\usepackage{amsfonts}

\newif\ifpdf
\ifx\pdfoutput\undefined
   \pdffalse
\else
   \pdfoutput=1
   \pdftrue
\fi

\ifpdf
    \usepackage[pdftex]{graphicx}
    \usepackage{color}
    \usepackage[pdftex,
               breaklinks=true,
               colorlinks=false,
               ]{hyperref}
\else
    \usepackage[dvips]{graphicx}
    \usepackage[hypertex,breaklinks=true,colorlinks=false]{hyperref}
\fi

\begin{document}
\newcommand{\Z}{\mathbb{Z}_2}

\title{Quantum  dimer model with $\Z$ liquid ground-state:
interpolation between cylinder and disk topologies and toy model
for a topological quantum-bit
}

\author{Gr\'egoire Misguich,  Vincent Pasquier}
\affiliation{Service de Physique Th\'eorique,
CEA-Saclay, 91191 Gif-sur-Yvette C\'edex, France}

\author{Fr\'ed\'eric  Mila}
\affiliation{Institut de th\'eorie des ph\'enom\`enes physiques,
  \'Ecole Polytechnique F\'ed\'erale de Lausanne BSP, \\ CH-1015
  Lausanne, Switzerland}

\author{Claire Lhuillier}
\affiliation{Laboratoire de Physique Th\'eorique des Liquides\\
Universit\'e P. et M. Curie and UMR 7600 of CNRS\\
Case 121, 4 Place Jussieu, 75252 Paris C\'edex, France}

\begin{abstract}
We  consider a quantum dimer model  (QDM)  on the kagome lattice which
was introduced  recently  [Phys.   Rev.    Lett.   {\bf 89},    137202
(2002)]. It realizes a $\Z$ liquid phase and its spectrum was obtained
exactly. It  displays a topological degeneracy when  the lattice has a
non-trivial geometry (cylinder, torus, etc).  We discuss and solve two
extensions   of the model    where    perturbations along lines    are
introduced:  first  the  introduction  of  a  potential  energy   term
repelling (or  attracting) the dimers along a  line is added, second a
perturbation  allowing to create, move  or destroy monomers.  For each
of these perturbations   we show that  there  exists a critical  value
above which,   in the  thermodynamic   limit, the   degeneracy  of the
ground-state is lifted from 2 (on a cylinder) to 1.  In both cases the
exact value of the gap  between the first two  levels is obtained by a
mapping to an Ising chain in transverse field.  This model provides an
example of  solvable Hamiltonian for  a  topological quantum bit where
the two perturbations act  as a diagonal and  a transverse operator in
the two-dimensional subspace.  We discuss how crossing the transitions
may be  used  in the   manipulation of  the  quantum  bit  to optimize
simultaneously the  frequency of operation   and  the losses  due   to
decoherence.

\end{abstract}
\maketitle

\section{Introduction}

Quantum  dimer     models   (QDM)\cite{rk88,ml03_5}  provide    simple
examples\cite{ms01,msp02}    of  microscopic    Hamiltonians      with
short-ranged resonating  valence-bond ground-states (or  dimer liquid)
with gapped excitations and no broken symmetry ($\Z$ liquids).  It has
been known  for  a long time   that such liquids are characterized  by
topological order:\cite{wen91} although the  system breaks no symmetry
and  has no  local    order parameter,  the ground-state  acquires   a
degeneracy (in the thermodynamic limit) which depends  on the genus of
the  surface where the    model  is defined (disk, cylinder,    torus,
etc.).  Remarkably, such  a topological  degeneracy is insensitive  to
small      local  perturbations   (such    as     weak    disorder for
instance).\cite{kitaev,ioffe02,iif02} On the other  hand, it  is clear
that   strong   enough   local   perturbations    should   lift   this
degeneracy.  Consider for instance a  $\Z$ dimer liquid on a cylinder,
with   a  two-fold degenerate  ground-state.  We  turn  on an external
potential  which penalizes   (with an  energy  $\lambda>0$)  any dimer
sitting across  a line extending from one  edge of the cylinder to the
other.    For very large $\lambda$,     one effectively ``cuts''   the
cylinder    down to   a disk    topology   and  one expects a   single
non-degenerate ground-state. It is therefore natural to expect a phase
transition at some intermediate value $\lambda$.

We  provide  here    a  simple model  where  the   spectrum,  and  the
ground-state degeneracy in particular, can be {\em exactly} calculated
as a function of $\lambda$ and the system size. This model generalizes
a QDM on the kagome lattice  (network of corner-sharing triangles with
triangular  and hexagonal plaquettes, see Fig.~\ref{fig:arrow})) which
was   introduced  recently.\cite{msp02}    The full   spectrum    (and
wave-functions) can  be obtained in  an elementary way and excitations
consist  of  static   and non-interacting   Ising  vortices\cite{rc89}
(visons\cite{sf01}).   In this paper we show   how the solution of the
model  extends to a situation where  an external  potential is applied
along a line  of the system.  The solution  is obtained by noting that
the bulk of the system decouples from the line and the line is exactly
described by an Ising chain in transverse field  (ICTF).  As a result,
we find a finite {\em critical} value  $\lambda_c$ of the perturbation
below which the  system behaves as  a cylinder (the  energy difference
between the two  quasi-degenerate ground-states is exponentially small
in the system size).  For $\lambda>\lambda_c$  the system behaves as a
disk and the ground-state is separated from the first excited state by
a finite gap $\mathcal{O}(\lambda-\lambda_c)$.

It  has been argued that  gapped systems with a topological degeneracy
could  provide physical ways  to implement quantum-bits (qubits) which
would  be   protected   from    decoherence  by    their   topological
nature.\cite{kitaev,ioffe02,ioffe02b,dfii04} Since   no   {\em  local}
measurement can distinguish the different  ground-states if the system
is  infinitely  large,    any  manipulation  (unitary    rotation)  or
measurement  (projection) of  the   state of the   qubit through local
observables will have to rely on finite-size effects.  We discuss this
issue in Section~\ref{sec:qubit} in  the light of the present solvable
model.  The  effective Hamiltonian acting on the  two lowest levels is
expressed in  terms of two generators $T^x$  and $T^z$ of rotations of
the qubit about   two quantization axis.  We  finally explain how  one
could take advantage of the phase transition at $\lambda=\lambda_c$ to
perform unitary rotations.   The  problems of this approach,  such  as
thermal excitations, will also be discussed briefly.

\section{Arrow representation, solvable QDM and topological degeneracy}

We consider  a QDM defined on  a kagome lattice with periodic boundary
conditions  in one direction (cylinder)  but the  arguments are easily
generalized to other topologies.

\subsection{Arrow representation}

Because  it is the  natural formalism  to describe and  solve the  QDM
discussed here,  we begin  by  reminding the representation   of dimer
coverings   of          the kagome lattice       in    terms   of {\em
arrows}.\cite{ez93,msp02}

The sites of a  kagome lattice $K$ (noted  $i$) can be identified with
bonds of the hexagonal  lattice $H$.\footnote{In fact  $H$ can be  any
trivalent lattice (each  site has three  neighbors) with any topology:
disk, sphere,  cylinder, torus etc...  As   an example, $H$ can  be an
hexagonal lattice with periodic boundary  conditions in one  direction
(cylinder).  Then    one constructs a new lattice    $K$ by the medial
lattice construction: sites of $K$ are the middle  points of the bonds
of $H$.\cite{medial} In the  following, for simplicity, we will choose
$H$ to be an hexagonal lattice and $K$ as a  kagome lattice (hence the
names $H$ and $K$).}   The triangles of $K$  (noted $t$) are sites  of
$H$.  As for hexagons  of $K$  (noted $h$),   they also correspond  to
hexagons of $H$.

\begin{figure}
  \begin{center}
    \includegraphics[width=3.5cm,angle=90]{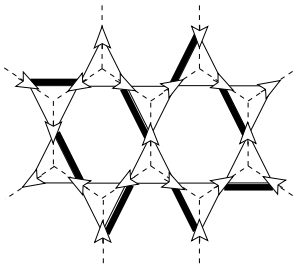}
    \caption[99]{
      A  dimer covering on  the kagome lattice
    (fat  bonds).    The corresponding   representation  with arrows
    living on the bonds of the hexagonal lattice (dashed lines)
    is displayed.}
    \label{fig:arrow}
  \end{center}
\end{figure}

From a (fully-packed) dimer covering of $K$  we orientate the bonds of
$H$ (arrows) in the following way: Each  bond of $H$  is a site of $K$
which  has one   dimer,  the corresponding   arrow points toward   the
interior of the triangle (of $K$) where the other end of the dimer is.
This is illustrated  in  Fig.~\ref{fig:arrow}.  As a  consequence, the
number of incoming  arrow(s) is even  (0 or 2) at  each vertex of $H$.
Inversely, any arrow configuration satisfying the parity constraint at
each vertex defines a unique dimer covering.

We can now define   the operators $\tau^x$, $\tau^z$ and
$\sigma^x$ acting  on the arrows.  The notations are those of
Refs.~\onlinecite{msp02} and \onlinecite{ml03_56}:

\begin{itemize}
    \item  $\tau^z_i$: flips the  arrow   at site  $i\in K$.   Any
        product $\tau^z_i\tau^z_j\cdots$ around  a {\em close loop} (of
        $H$) is a ``physical'' operator in the sense that it conserves
        all the constraints.

        \item $\sigma^x_h=\prod_{i=1}^6 \tau^z_i$.  Flips the 6 arrows
        $i=1\cdots6$ around the  hexagon $h$ (smallest closed  loop on
        $H$).\footnote{In terms of  dimers, $\sigma^x_h$ is the sum of
        all possible  ``kinetic  energy'' terms  (ring exchange) which
        can be defined inside the   star surrounding $h$. Those  terms
        allow  from 3  to 6  dimers to move.}   From the  definition of
        $\tau^z$   above,    the  $\sigma^x$     operators     satisfy
        $(\sigma^x)^2=1$ and commute with each other.

        \item $\tau^x_i$: Compares the arrow at  site $i$ with the the
        arrow  in some (arbitrary) reference   covering ($=+1$ if they
        are  the same, $-1$   otherwise). The hard-core constraint  on
        dimers translates  into $\tau^x_0\tau^x_1\tau^x_2=1$ for every
        triangle $(012)$ of the kagome lattice.
\end{itemize}

From now on and for most purposes one can forget the dimers themselves
and   focus  only    on    the      bond    degrees  of        freedom
$\tau^x_i=\pm1$.\footnote{Those  bond variables are  $\Z$ gauge degrees
of freedom\cite{msp02} while the local constraint is  the Gauss law of
the  gauge description.\cite{msf02}   As for  $\sigma^x_h=\prod_{i=1}^6
\tau^z_i$, it     is the  gauge   flux   going  through  the   hexagon
$h$.}  We note  that,   in principle, the $\tau^x_i=\pm1$   degrees of
freedom could  be physically realized   with  real spins  living on  a
kagome lattice. A strong easy-axis anisotropy could then force them to
point toward the center of one of the neighboring triangle.

\subsection{Bulk Hamiltonian}

The QDM introduced in Ref.~\onlinecite{msp02} is
\begin{equation}
   \mathcal{H}_0=-\sum_h \sigma^x_h=-\sum_h \prod_1^6 \tau^z_{h_i}
   \label{eq:h0}
\end{equation}
All   the eigenstates are  easily   obtained because the  $\sigma^x_h$
operators  commute    with  each   other  and  have   two  eigenvalues
$\sigma^x_h=\pm1$.\footnote{They     are   independent     pseudo-spin
operators.  Notice  however that if the  system  has no edge (torus or
sphere), they are subjected to  the an additional constraint  $\prod_h
\sigma^x(h)=1$. This  comes from the  fact  that, in such  a geometry,
$\prod_h\sigma^x(h)$ flips every arrow {\em  twice} and thus reduces to
the identity.}

\subsection{Topological sectors}

As usual     for  dimer models, the   configurations    are grouped in
topological sectors (TS): two configurations are in the same TS if and
only if  they can be transformed  into  each other  by a succession of
{\em local}\footnote{Here local means   that the associated loop  does
not  wind around the whole  system (cylinder).}   moves.  As explained
below, there are two TS when the system has the topology of a cylinder.

\begin{figure}
  \begin{center}
    \includegraphics[width=5cm]{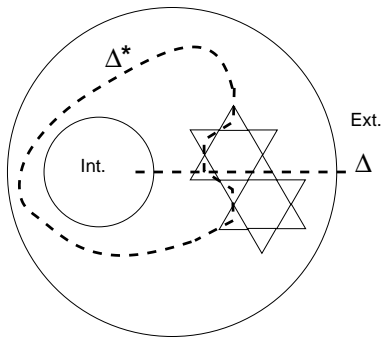}
    \caption[99]{
      Kagome lattice on a cylinder  with a cut $\Delta$  going from one edge
      of the cylinder  to the other.  The dual cut $\Delta^*$ passes through
      the centers of the triangles and winds around the cylinder.}
    \label{fig:cyl}
  \end{center}
\end{figure}

\begin{figure}
  \begin{center} \includegraphics[width=8.3cm]{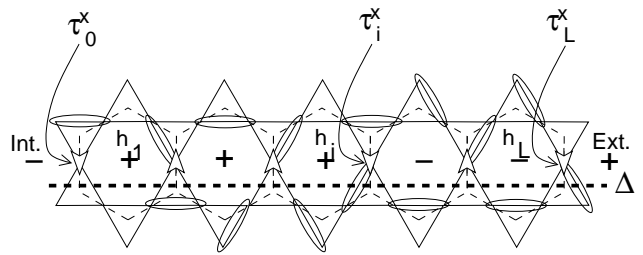}

\caption[99]{
  Dimer covering  of  the kagome lattice  in  the vicinity  of the cut
  $\Delta$  (dashed line). The arrows next   to $\Delta$ (appearing in
  Eq.~\ref{eq:n})     are  shown.   The     signs   in  the   hexagons
  Int.,$h_1,\cdots, h_L$,Ext. indicate the values of the corresponding
  pseudospins  $\sigma^z$ (with   the  assumption that  the  reference
  configuration has no dimer crossing $\Delta$).}

\label{fig:chain} \end{center}
\end{figure}

First draw a  cut $\Delta$ (crossing the  bonds of the lattice)  going
from one edge of the cylinder to the  other (Fig.~\ref{fig:cyl}).  Let
$N_\Delta$ be the  number of dimers crossing  $\Delta$.  It has  a
simple expression in terms of the $\tau^x_i$:
\begin{eqnarray}
        N_\Delta=\frac{1}{2}\sum_{i=0}^L (1-\tau^x_i)
        \label{eq:n}
\end{eqnarray}
where  the sites $i=0\cdots L$ are  the  centers of  the  bonds of $H$
which  are  cut by $\Delta$,  as  shown  in Fig.~\ref{fig:chain}.  For
simplicity,   we assumed in   Eq.~\ref{eq:n}  that  no  dimer  crosses
$\Delta$ in the reference covering.

Any {\em local} dimer move  conserves the parity  of $N_\Delta$ and it
is natural to define:
\begin{equation}
        T^x=\prod_{i\in\Delta}\tau^x_i=(-1)^{N_\Delta}
        \label{eq:Tx}
\end{equation}
All coverings  with $T^x=1$ (resp.   $-1$) define a  TS, called  $S_+$
(resp. $S_-$)  and $\mathcal{H}_0$ can  be  diagonalized separately in
each sector.

\subsection{Topological degeneracy of $\mathcal{H}_0$}
It is straightforward to check that $\mathcal{H}_0$ has the same
energy in each sector. Let $\Delta^*$   be a closed loop
encircling   the cylinder (Fig.~\ref{fig:cyl}) and define an
operator
\begin{equation}
        T^z=\prod_{i\in\Delta^*}\tau^z_i
        \label{eq:Tz}
\end{equation}
which flips all the corresponding arrows. $T^z$  commutes with all the
$\sigma^x$ operators and maps $S_+$ onto $S_-$:
\begin{equation}
    T^zT^x=-T^xT^z
\end{equation}
This shows that if $|\psi\rangle$ is an eigenstate of $\mathcal{H}_0$,
$T^z|\psi\rangle$ is an  eigenstate  of $\mathcal{H}_0$ with the  same
energy (but  in   the other   TS).   This demonstrates  the   two-fold
(topological) degeneracy of the eigenstates of $\mathcal{H}_0$.

\section{Perturbation lifting the degeneracy between the $T^x=\pm1$ sectors}

We  introduce a  potential energy term   which couples  to the  dimers
crossing $\Delta$:\footnote{The case where the potential term pins the
dimers along some reference configuration in the {\em whole} system is
discussed in Ref.~\onlinecite{msp02}  and is equivalent  to a 2D Ising
model in transverse field.}
\begin{eqnarray}
        \mathcal{H}_1&=&2 \lambda N_\Delta \\
                     &=&\lambda \sum_{i=0}^L (1-\tau^x_i)
        \label{eq:h1}
\end{eqnarray}
As discussed  in the introduction,  we expect  that  a small $\lambda$
should not affect the two-fold  degeneracy while $\lambda\gg 1$ should
leave      a      single     ground-state.       In    presence     of
$\mathcal{H}=\mathcal{H}_0+\mathcal{H}_1$,  $T^x=(-1)^{N_\Delta}$   is
still   a  conserved  quantity  but   $T^z$   does not  commute   with
$\mathcal{H}_1$   and the two 
sectors
are   no  longer degenerate   when
$\lambda\ne0$. 
Since for $\lambda\gg1$ the system minimizes $N_\Delta$,
the ground-state of the $T^x=+1$ sector tends to a state with $N_\Delta\simeq0$
and that of the  $T^x=-1$ sector to a state  with $N_\Delta\simeq1$.
Instead of having a superposition of dimer
configurations with different values of $N_\Delta$ but a fixed {\em  parity}
({\em non-local} observable), the large $\lambda$ limit corresponds to a well
defined $N_\Delta$ (sum of {\em local} operators).
While $T^x$ is still a conserved quantity, 
we already see that the topological nature of the $T^x=+1$ and $T^x=-1$
sectors disappears when $\lambda$ is large. 

The  perturbation $\mathcal{H}_1$ is identical  to the
one introduced  by    Ioffe   {\it  et al.}~\cite{ioffe02}     in    a
triangular-lattice QDM   in  order to  manipulate  (``phase shifter'')
their qubit.  However, in  our case, {\em  the  existence of  an arrow
representation makes it possible to calculate exactly the spectrum} of
$\mathcal{H}=\mathcal{H}_0+\mathcal{H}_1$.

\subsection{Mapping to the ICTF}\label{sec:cylinder}

The  Hamiltonian $\mathcal{H}=\mathcal{H}_0+\mathcal{H}_1$
(Eqs.~\ref{eq:h0} and \ref{eq:h1}) can be separated into a
``bulk'' and a ``chain'' part as below :
\begin{eqnarray}
        \mathcal{H}&=&\mathcal{H}_\Delta+\mathcal{H}_{\rm bulk}
        \label{eq:hdb}\\
        \mathcal{H}_\Delta&=&-\sum_{i=1}^L\sigma^x_{h_i}
                +\lambda\sum_{i=0}^L (1-\tau^x_i)       \\
        \mathcal{H}_{\rm bulk}&=&-\sum_{h'\notin\Delta} \sigma^x_{h'}
\end{eqnarray}
It    is  important  to    emphasize  that  $\mathcal{H}_\Delta$   and
$\mathcal{H}_{\rm   bulk}$ {\em commute  with    each other} and   can
therefore  be treated   separately.  From  now   on we concentrate  on
$\mathcal{H}_\Delta$.

\begin{figure}
  \begin{center}
    \includegraphics[width=4cm]{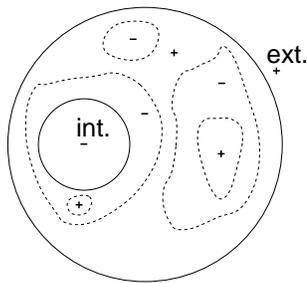}
    \caption[99]{ Cylinder  (full lines) and loops  (dashed lines)
      along  which   a  dimer  configuration $c$  differs   from the
      reference.  The signs indicate the value of $\sigma^z$ in each
      domain.}
    \label{fig:sz}
  \end{center}
\end{figure}

{\em  $\sigma^z$ pseudo-spins}.---   A  $\sigma^z_h$  operator can  be
introduced for each hexagon $h$ in the following way. Due to the local
constraint ($\tau^x_i\tau^x_j\tau^x_k=1$ on each triangle of $K$), the
bonds   of $H$ where    $\tau^x=-1$ necessarily  form non-intersecting
closed loops.\footnote{These are bonds where the arrows have different
orientations in  $c$ and in the reference.}   We interpret these loops
as domain walls for   some Ising pseudo-spins  $\sigma^z_h=\pm1$ which
leave on each  hexagon.  To  remove the  two-fold ambiguity  we assume
that the exterior has  a fixed spin  $\sigma^z_{\rm ext}=1$.  In turn,
this defines a  $\sigma^z_{\rm int}$  associated to  the  ``interior''
(Fig.~\ref{fig:sz}).  It is easy to check that  this Ising spin labels
the     TS    of        the  configuration      because $\sigma^z_{\rm
int}=\prod_{i\in\Delta^*}\tau^x_i=T^x$.       The       other ``bulk''
pseudo-spins  are   those  introduced by  Zeng  and  Elser.\cite{ez93}
Eventually we mention  that $\sigma^z_h$ and $\sigma^x_h$  anticommute
(they commute if not  on the same  hexagon), as suggested by the Pauli
matrix  notation. This  is  easily   checked from  the  definition  of
$\sigma^x_h$ in terms of arrows.
From    the  definition        of   $\sigma^z_h$   we      have    the
relation~:\cite{msp02}$^,$\footnote{The relation   between the $\tau$  and
$\sigma$ operators  is the same  as the standard duality between Ising
models    and $\mathbb{Z}_2$ gauge  theory  in   2+1 dimensions.   See
Ref.~\onlinecite{msp02}}
\begin{equation}
        \sigma^z_{h} \sigma^z_{h'}=\tau^x_i
        \label{eq:sst}
\end{equation}
where $h$ and $h'$ are the two hexagons  touching $i$. For an arrow
$i_0$ next to a boundary (say interior), this relation is modified
to:
\begin{equation}
        \sigma^z_{h} \sigma^z_{\rm int}=\tau^x_{i_0}
        \label{eq:sst2}
\end{equation}
These relations are used to get:
\begin{equation}
        \mathcal{H}_\Delta=
        -\sum_{i=1}^{L}\sigma^x_{h_i}
        - \lambda \sum_{i=0}^{L} \sigma^z_{h_i} \sigma^z_{h_{i+1}}
        + (L+1)\lambda
        \label{eq:chain}
\end{equation}
where       the   two    boundary       spins    are identified     to
$\sigma^z_0=\sigma^z_{\rm int}$  and $\sigma^z_{h_{L+1}}=\sigma^z_{\rm
ext}$.    It  appears that   $\mathcal{H}_\Delta$  is nothing  but the
Hamiltonian of an open ICTF  with a magnetic  exchange $\lambda$ and a
transverse field equal to 1.   This model can be  solved by a standard
Jordan-Wigner transformation and  maps  onto  free fermions.   In  the
thermodynamic  limit      it   has     a   paramagnetic    phase  with
$\langle\sigma^z_h\rangle=0$    for   $|\lambda|<\lambda_c=1$   and an
ordered     phase   with      $\langle\sigma^z_h\rangle\ne0$       for
$|\lambda|>\lambda_c=1$.

The boundary  spins   $\sigma^z_0$ and   $\sigma^z_{h_{L+1}}$  play  a
special role.  $\sigma^z_{h_{L+1}}=\sigma^z_{\rm ext}$ is fixed to $1$
but  $\sigma^z_0=\sigma^z_{\rm int}$  is    free  (but  conserved   by
$\mathcal{H}_\Delta$) and labels the TS.  The spectrum of the ICTF can
thus be studied separately for $\sigma^z_0=+1$ or $\sigma^z_0=-1$. One
takes  $\lambda\geq0$     (ferromagnetic    chain)  without    loss of
generality.  The sector with $\sigma^z_0=+1$  thus corresponds to {\em
unfrustrated boundary conditions} for the (pseudo-spin) chain.  On the
other hand, choosing  $\sigma^z_0=-1$ amounts to  impose  at least one
Ising    domain  wall  in the    system.\footnote{The relation between
topological sectors of QDM and  boundary conditions in spin models was
already  noted  in  Refs.~\cite{msp02,msf02,ms03},  together with  the
connexion between confinement transition and Ising transition.} In the
thermodynamic limit both  sectors have the same  energy {\em per site}
but they may have a  finite difference in the total  energy (gap).  In
the paramagnetic phase  ($\lambda<1$), because of the finite spin-spin
correlation   length    $\xi(\lambda)$,    the  frustration   has   an
exponentially small  effect on the  ground-state energy and the energy
difference between  the  two TS  is   $\Delta E\sim\exp(-L/\xi)$  with
$\xi\simeq   \ln(1/\lambda)^{-1}$ (see Ref.~\onlinecite{dfii04}    and
Appendix~\ref{sec:eictfO}).  On the other hand,
in the ferromagnetic phase   ($\lambda>1$),  the Ising spins  want  to
order and  the boundary condition $\sigma^z_0\ne\sigma^z_L $ generates
a  {\em  finite  energy}  cost   $\Delta E\sim\mathcal{O}(L^0)$   (see
Eq.~\ref{eq:delta1}).

The   result    is thus  that  for    $\lambda<1$  the ground-state is
asymptotically two-fold degenerate   in the thermodynamic limit.   The
gap between the  two  TS is $\Delta E\sim\exp(-L/\xi(\lambda))$  where
$\xi$ is  the correlation length  of  the ICTF.  In  the thermodynamic
limit there  is a finite critical  value $\lambda_c=1$ above which the
topological  degeneracy   is
destroyed.
Above  $\lambda_c$   the Ising
pseudo-spins have a  positive magnetization.  Since $\sigma^z_h$ is an
operator  which creates  an Ising  vortex  (vison\cite{msp02}), it  is
natural to  interpret $\langle\sigma^z_h\rangle>0$ as the existence of
a  {\em condensate}   of  those  particles  (along $\Delta$).     This
condensation is responsible for changing the ``effective'' topology of
the system from  a cylinder into a  disk.  In  this simple model  what
happens along the  chain $\Delta$  is decoupled from  the bulk  of the
system.  The  perturbation caused by the  potential $\lambda$ does not
extend into the bulk, which  remains a liquid with non-interacting and
static visons excitations.

\section{Mixing the $T^x=\pm1$ sectors with monomers}

The perturbation $\mathcal{H}_1$ discussed  in the previous section is
not the only way to
remove
the topological degeneracy.  It is well known
that in the presence of mobile monomers, $T^x=(-1)^{N_{\Delta}}$ is no
longer conserved.\footnote{If a monomer winds  around the cylinder  it
connects a configuration $T^x=1$  with a configuration $T^x=-1$.} This
property was    used in   Refs.~\cite{ioffe02,ioffe02b}  to   mix  the
topological  sectors.   We will show that   in our kagome geometry the
arrow representation of the dimer model  allows to compute exactly the
spectrum of the system when monomers are allowed to be created, to hop
and to be  destroyed along a  line winding around  the cylinder. As we
will  see  this model is   closely related to  that  discussed  in the
previous  section: $\Delta$ is replaced   by $\Delta^*$, monomers will
play the role   of the visons and    the ICTF will have  periodic  and
antiperiodic boundary conditions instead of open ones.

\subsection{Hamiltonian}
\begin{figure}
\begin{center}
\includegraphics[width=5.5cm]{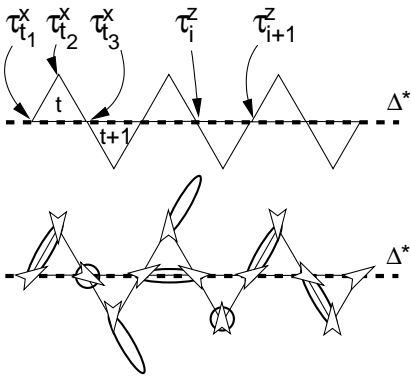}
\caption[99]{Top: Kagome lattice  in  the vicinity  of the cut
  $\Delta^*$ (dashed line).  Bottom: Mixed dimer-monomer configuration
  and the arrow representation.}
\label{fig:monomers}
\end{center}
\end{figure}

We             relax         the             parity         constraint
$\tau^x_{t_0}\tau^x_{t_1}\tau^x_{t_2}=1$ on each  triangle $t$ so that
triangles with one   or three incoming  arrows  are allowed.  When   a
triangle has {\em one} incoming arrow, it  is naturally interpreted as
the presence of a monomer  (or hole) at the   site of this arrow  (see
Fig.~\ref{fig:monomers}). When  a  triangle has {\em   three} incoming
arrows, we interpret it as a monomer and a dimer which are delocalized
over  the three sites.\footnote{In  the $\Z$ gauge theory language the
constraint is  the    Gauss    law   and allowing     triangles   with
$\tau^x_{t_0}\tau^x_{t_1}\tau^x_{t_2}=-1$     amounts to allow   gauge
charges  (``matter'') in  the system.}   Flipping  one arrow ({\it  i.e}
acting  with    $\tau^z_i$) on a  dimer   state  therefore creates two
monomers on the nearby triangles (one of which may be delocalized over
three sites). We consider the following Hamiltonian:
\begin{eqnarray}
        \mathcal{H}_0'
                &=&-\sum_h \sigma^x_h
                        +U\sum_t (1-\tau^x_{t_0}\tau^x_{t_1}\tau^x_{t_2}) \\
                &=&-\sum_h \prod_{i=1}^6 \tau^z_{h_i}
                        +U\sum_t (1-\tau^x_{t_0}\tau^x_{t_1}\tau^x_{t_2})
        \label{eq:hbulk}
\end{eqnarray}
where $h_{1\cdots6}$ are the  sites around hexagon $h$ and $t_{1,2,3}$
the sites of the triangle  $t$ (see Fig.~\ref{fig:monomers}). $U$ is a
large energy enforcing the constraint on low-energy states.
This type of model  was  first considered by Kitaev.\cite{kitaev}  For
$U>0$ the ground-state  of  the Hamiltonian Eq.~\ref{eq:hbulk}  is the
same  as    for  $U=\infty$  (equivalent   to   Eq.~\ref{eq:h0}) since
$\tau^x_{t_0}\tau^x_{t_1}\tau^x_{t_2}$    commutes    with   all   the
$\sigma^x$. However,  static pairs of  monomers are present in excited
states with energies greater than $2U$ above the ground-state.

As Ioffe~{\it et  al.},\cite{ioffe02,ioffe02b} we  wish to use   these
monomers to  couple  the two TS.  For  this purpose the   monomers are
allowed   to  be created   and  to  propagate   along one  closed loop
$\Delta^*$   winding  around  the  cylinder  (Fig.~\ref{fig:cyl}). The
simplest term which does this is
\begin{eqnarray}
  \mathcal{H}_1'
  &=&-\mu U\sum_{i\in\Delta^*}\tau^z_i
  \label{eq:hd*}
\end{eqnarray}
(this choice of normalization will make the analogy with the $\lambda$
perturbation clearer). When a $\tau^z_i$ term acts on a site located
between     two   triangles          satisfying   the       constraint
($\tau^x_{t_0}\tau^x_{t_1}\tau^x_{t_2}=1$),  it  creates  a   pair  of
monomers.    When it  acts  on   a pair  of  triangles  violating  the
constraint, the pair of monomers  is destroyed.  If it  acts on a site
located    between    triangles       with different      values    of
$\tau^x_{t_0}\tau^x_{t_1}\tau^x_{t_2}$,     a  monomer  hops from  one
triangle to the other.
\subsection{Mapping to the ICTF}
\begin{figure}
  \begin{center} \includegraphics[width=6cm]{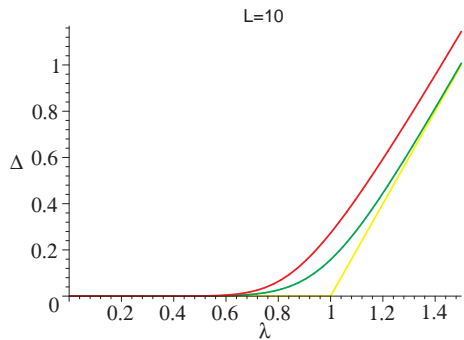} \caption[99]{ Effect
  of a change of boundary conditions on the  ground-state energy of an
  ICTF with $N=10$ spins as a function of the (ferromagnetic) exchange
  $\lambda$.   Circles: closed chain  with periodic  and anti-periodic
  boundary conditions.   Thin  line:   Ground-state energy  difference
  between open chain with ferromagnetic ($\uparrow\cdots\uparrow$) and
  antiferromagnetic ($\downarrow\cdots\uparrow$) boundary  conditions.
  Thick  line: $N=\infty$ case (same result  for   the open and  closed
  chains). } \label{fig:gap} \end{center}
\end{figure}

As in Eq.~\ref{eq:hdb}, $\mathcal{H}_0'+\mathcal{H}_1'$ splits
into a bulk  and  a  one-dimensional parts:\footnote{$U$ needs not
have the same value in the bulk and along $\Delta^*$. One can also
choose $U=\infty$ in the bulk  and $U<\infty$ along $\Delta^*$ so
that monomers can exist and propagate only along the the chain.}
\begin{eqnarray}
  \mathcal{H}_0'+\mathcal{H}_1'&=&\mathcal{H}_{\rm bulk}'+\mathcal{H}_{\Delta^*}
  \label{eq:hdb'}\\
  \mathcal{H}_{\rm bulk}'&=&-\sum_{h} \sigma^x_{h}
  +U\sum_{t\notin\Delta^*} (1-\tau^x_{t_0}\tau^x_{t_1}\tau^x_{t_2})\\
  \mathcal{H}_{\Delta^*}&=&-\mu U\sum_{i\in\Delta^*}\tau^z_i
  +U\sum_{t\in\Delta^*} (1-\tau^x_{t_0}\tau^x_{t_1}\tau^x_{t_2})\hspace*{0.3cm}
\end{eqnarray}
One    can   simply     check    that  $\mathcal{H}_{\Delta^*}$    and
$\mathcal{H}_{\rm  bulk}'$ commute with each  other so that one has to
study a one-dimensional model $\mathcal{H}_{\Delta^*}$.  This model is
identical  to  a closed ICTF  (with  periodic or antiperiodic boundary
conditions) as  explained below.  Each {\em  triangle} $t$  crossed by
$\Delta^*$  corresponds  to a   {\em   site} of   the spin  chain. The
associated transverse-field term for this Ising spin is:
\begin{equation}
    \tilde\sigma^x_t=\tau^x_{t_0}\tau^x_{t_1}\tau^x_{t_2}
\end{equation}
We define the $z$ component of the spins as
\begin{eqnarray}
    \tilde\sigma^z_t=\tau^z_0 \tau^z_1 \cdots ... \tau^z_{i(t)}
\end{eqnarray}
where  $0$ is an  (arbitrary) origin  on $\Delta^*$  and $i(t)$ is the
site of $K$ in common with  triangles $t$ and $t-1$.   It is simple to
check  that  the $\tilde\sigma^x$  and $\tilde\sigma^z$  defined above
(not to be confused  with  $\sigma^x$ and  $\sigma^z$) obey the  usual
Pauli   matrix algebra  and    play the   role of   spin-$\frac{1}{2}$
operators.  In addition, these definitions insure that
\begin{equation}
    \tilde\sigma^z_t\tilde\sigma^z_{t+1}=\tau^z_{t_3}
    \label{eq:sigmatz}
\end{equation}
where $t_3$ is the common site between triangles $t$  and $t+1$ (as in
Fig.~\ref{fig:monomers}). Because $\Delta^*$  is  a  closed curve,   a
special care is needed for the last term:
\begin{equation}
    T^z \;\tilde\sigma^z_{L-1}\tilde\sigma^z_0=\tau^z_0
    \label{eq:sigmat-z}
\end{equation}
where  $T^z$      (defined     in  Eq.~\ref{eq:Tz})    commutes   with
$\mathcal{H}_{\Delta^*}$      (in  the  same      way  we   had before
$[\mathcal{H}_\Delta,T^x]=0$). It can be successively set to $\pm1$ to
obtain      the    whole     spectrum.      With    these    notations
$\mathcal{H}_{\Delta^*}$ reads
\begin{eqnarray}
    \frac{1}{U}\mathcal{H}_{\Delta^*}&=&
    -\mu \left(
    T^z\sigma^z_{L-1}\tilde\sigma^z_0
    +\sum_{t=0}^{L-2}\tilde\sigma^z_t\tilde\sigma^z_{t+1}
    \right) \nonumber\\
    &&-\sum_{t=0}^{L-1}\tilde\sigma^x_t + {\rm cst.}
\end{eqnarray}
This is the Hamiltonian  of an ICTF  with periodic boundary conditions
when $T^z=1$ and antiperiodic boundary  conditions when $T^z=-1$.   As
for $\mathcal{H}_{\Delta}$, the model has, in the thermodynamic limit,
a phase transition at  $\mu_c=1$ between  a  paramagnetic phase and  a
ferromagnetic phase.   The  calculation  of  the gap between   the two
sectors amounts to study a  closed ICTF with periodic and antiperiodic
boundary conditions.   The  exact  result  for  the energy  difference
$\Delta E'$ is derived in the Appendix
\ref{sec:eictfC} (see also  Ref.~\onlinecite{dfii04}) with the help of
a Jordan-Wigner transformation. In the limit  of a large system ($L\gg
1$) it is given by
\begin{eqnarray}
        \Delta E'&=&E_+-E_-\nonumber \\
    &\simeq& 2U\sqrt{\frac{1-\mu^2}{L\pi}}\mu^{L}
        \;\;{\rm for}\; \mu<1
        \label{eq:mu<1} \\
        &\simeq& 2U(\mu-1)
        \;\;\;\;\;\;\;\;\;\;{\rm for}\; \mu>1
        \label{eq:mu>1}
\end{eqnarray}
The  result  for  $L=10$ is plotted   in Fig.~\ref{fig:gap}.    As for
$\mathcal{H}_{\Delta}$, the   critical  value  $\mu_c=1$ separates   a
regime with  an  exponentially  small  splitting between  the  sectors
(Eq.~\ref{eq:mu<1}) and  a  regime with   a finite gap  between   them
(Eq.~\ref{eq:mu>1}). In the spin language,  the ferromagnetic phase is
characterized by  $\langle\tilde\sigma^z\rangle\ne0$.   Going  back to
dimer/monomers variables,  we  find that $\tilde\sigma^z_t$  flips all
the arrows located on $\Delta^*$ between the origin and $t$.  Thus, it
creates or annihilates a pair of monomers  sitting at both ends of the
string or  move a monomer from   one end to the  other.  In both cases
$\tilde\sigma^z_t$ creates  or   destroys a monomer   on  the triangle
$t$.\footnote{This should  be compared with $\sigma^z_h$ which creates
a vison on  hexagon $h$.}  $\langle\tilde\sigma^z\rangle\ne0$ can thus
be interpreted as a  {\em condensation} of monomers  along $\Delta^*$.
This is  equivalent to the condensate  of visons mentioned in the case
of $\mathcal{H}_\Delta$ when $\lambda>1$.

To conclude this section we discuss a difference between the $\lambda$
and the $\mu$ perturbations.  In the limit where $\lambda\to\infty$ no
dimer can  sit on  $\Delta$  anymore and  the torus  is  reduced  to a
rectangle, as if the lattice  had been cut  with scissors. If the same
geometrical picture was  true   for   the perturbation  $\mu$    along
$\Delta^*$,  one  would   erroneously  conclude  that the   system  is
effectively     transformed  into   {\em     two}   cylinders with   a
$2\times2=4$-fold ground-state  degeneracy. This is  incorrect for the
following     reason.  The     path    $\Delta^*$  chosen  to   define
$T^z(\Delta^*)$ (Eq.~\ref{eq:Tz})  may  be  shifted  by multiplication
with a $\sigma^x$ operator:
\begin{equation}
	\sigma^x(h) T^z(\Delta^*_1) =  T^z(\Delta^*_2)
	\label{eq:TzTz}
\end{equation}
where $h$  is an  hexagon  next to  $\Delta^*_1$  and $\Delta^*_2$  is
identical  to $\Delta^*_1$ except that it  passes on the other side of
$h$.  From  Eq.~\ref{eq:TzTz}, we  see that $T^z(\Delta^*_1)|0\rangle$
does not depend on the location of $\Delta^*_1$ as long as this closed
curve is deformed by passing only on hexagons with $\sigma^x(h)=1$. In
the case of   a the perturbation   $\mu$, the  ground-state  satisfies
$\sigma^x(h)|0\rangle=|0\rangle$   for  all hexagons  $h$,  even those
along  $\Delta^*$ (this  is  not true  for the  $\lambda$ perturbation
which does   not  commute  with  $\sigma^x$).  Therefore   one  cannot
independently flip the  topological sector  of the ``upper''  cylinder
with some  $T^z_1$ without changing the sector  of  the ``lower part''
which     is        controlled       by    a        $T^z_2$,     since
$T^z_1|0\rangle=T^z_2|0\rangle$.

\section{A toy model for a topological qubit}
\label{sec:qubit}

Kitaev\cite{kitaev}   suggested    that  systems  with   topologically
degenerate   ground-states could be   used to realize qubits protected
from decoherence.  This suggestion was then made more precise by Ioffe
{\it   et   al.}\cite{ioffe02,ioffe02b} and     Dou{\c    c}ot{\it  et
al.}\cite{dfi03,dfii04} who proposed to use Josephson junctions arrays
to implement such  a  system.  As mentioned  in the  introduction, the
topological  nature  of   the   degeneracy   makes it  difficult    to
``manipulate''  (perform  unitary  rotation)   because  it is   almost
insensitive to  local  couplings.  On    the  other hand, it   may  be
difficult   to  apply  a  perturbation  corresponding  to  a non-local
operator (such as $T^z$ or  $T^y$).  If however  the system allows for
an hardware  implementation of  such a non-local operator,\footnote{In
the circuit  proposed in  Ref.~\onlinecite{ioffe02}, the state  of the
qubit could be measured  through a weak Josephson junction  connecting
the  tow edges  of the cylinder.}   it  represents a dangerous channel
through  which perturbations  could  bypass the topological protection
and  contribute   to decoherence of  the  qubit.   If such a non-local
coupling to  the system exists, one  must be able to ``disconnect'' it
efficiently when it is not active.

The   clever   solution proposed in Ref.~\onlinecite{ioffe02,ioffe02b}
consists in perturbing  the system by  two  external potentials acting
along two   lines,  exactly as $\Delta$    and $\Delta^*$. While these
perturbations are local (more precisely they are sums of local terms),
they  induce a  splitting of   the  two ground-states  proportional to
$\lambda^L$  and therefore induce a slow  precession of  the qubit. In
this section we take advantage of  the exact solution  of the model to
discuss these effects beyond the regime where $\lambda^L\ll1$.

\subsection{Unitary rotations}
Combining the perturbations along $\Delta$ and $\Delta^*$ the Hamiltonian is
\begin{eqnarray}
  \mathcal{H}(\lambda,\mu)&=&\mathcal{H}_{0'}+\mathcal{H}_{1'}+\mathcal{H}_{1}\\
  &=&-\sum_h \prod_{i=1}^6 \tau^z_{h_i}+U\sum_t (1-\tau^x_0\tau^x_1\tau^x_2)
  \nonumber\\
  &&-\mu U\sum_{i\in\Delta^*}\tau^z_i-\lambda\sum_{i\in\Delta}(\tau^x_i-1)
\end{eqnarray}
From     the      previous     calculations      we       know    that
$\mathcal{H}(\lambda,\mu=0)$  lifts the degeneracy   of the two TS. It
acts in this two-dimensional subspace as $\Delta E(\lambda)T^x$, where
$\Delta    E(\lambda)$,    given      in  Eqs.~\ref{eq:delta1}     and
\ref{eq:delta2}, is the  energy  difference between  ferromagnetic and
antiferromagnetic boundary   conditions  for  an   ICTF with  exchange
$\lambda$    (and  a  unit  transverse  field).    On  the other hand,
$\mathcal{H}(\lambda=0,\mu)$ mixes the   two sectors.   Its action  is
described     by  $U\Delta      E'(\mu)T^z$     where  $\Delta     E'$
(Eq.~\ref{eq:delta3})   is   energy difference   between  periodic and
antiperiodic boundary  conditions for an ICTF   (with exchange $\mu U$
and transverse field $U$).\footnote{Since any unitary operation can be
performed using  $T^x$  and $T^z$  sequentially,  there is no  need to
analyze the spectrum with $\lambda\ne0$ and $\mu\ne0$ simultaneously.}

If the  system is operated below the  critical values of $\lambda$ and
$\mu$, the qubit precesses at a frequency which is exponentially small
in  the  system size.  The  cylinder topology  and  the low density of
monomers protect the degeneracy of  the spectrum.  This is the  regime
mentioned by  Ioffe {\it et al.}.   However, the system size cannot be
too  large  because the  time required  for a   unitary rotation would
become exponentially long.  On  the other hand, if  the system  is not
large enough, its topology does not  fully protect it from decoherence
by external perturbations.

If one of the external parameters ($\lambda$ or $\mu$) is pushed above
its critical   value, the frequency  becomes finite,  even  in case of
large  system size.  We  may therefore   take advantage  of the  phase
transitions in   a large   system.  In   such  a   case the   qubit is
topologically  protected as  long as  $\lambda$ and  $\mu$ are smaller
than their critical values, even if they are not precisely set to zero
or if they introduce some noise. It is only during the ``manipulation"
($\lambda(t)>1$ or  $\mu(t)>1$) that  the state  of the  qubit evolves
(and is  sensitive to the  external noise entering through $\Delta$ or
$\Delta^*$).

To preserve an adiabatic evolution one must avoid transitions to other
eigenstates. However the gap in the spectrum of the ICTF becomes small
(of the order of $\sim 1/L$) in the vicinity  of the transition.  This
limits the typical time of the unitary rotation to be  at least of the
order of $L$.   This {\em linear} dependence in  the system size is an
interesting  property  because  it  should be   compared to  the  {\em
exponential}   dependence   ($\sim   \lambda^{-L}$) present    in  the
perturbative regime.  Also because  of this small gap, the temperature
must be  $T\ll  1/L$ to avoid  thermal excitations  when $\lambda$ (or
$\mu$) is close to  $1$.    Using the  transition to  perform  unitary
rotation therefore  seems to improve the time  of  operation and could
enable   to  use a  larger  system  and   to  benefit from  a stronger
topological protection.  It also requires to work at lower temperature
than for a    qubit operated in the  perturbative  ($\lambda,\mu\ll1$)
regime only and this may represent a severe limitation.

\subsection{Reading out the state of the qubit}
We assume that the   qubit  is in  a linear   combination of the   two
topological  sectors:  $|\psi\rangle=\alpha |+\rangle+\beta |-\rangle$
where $T^x|+\rangle=|+\rangle$ and $T^x|-\rangle=-|-\rangle$.  We wish
to measure $|\alpha|^2$  with  a {\em  local} observable.  This is not
directly  possible  if  the system    is very  large since any   local
observable has expectation values in $|+\rangle$ and $|-\rangle$ which
are exponentially close.  Likewise, a local observable has a vanishing
matrix  element   between $|+\rangle$ and   $|-\rangle$.   A  possible
procedure  could be to   switch adiabatically the  potential $\lambda$
above  the transition.   For a   strong  enough  $\lambda$, the  state
$|+\rangle$  evolves  to a  superposition  of dimer  coverings with no
dimer crossing $\Delta$.  On the other hand,  $|-\rangle$ evolves to a
superposition of  dimer    coverings with  {\em  one}  dimer  crossing
$\Delta$. This  is because the parity $(T^x)$  is a conserved quantity
under the evolution but the ground-state has to minimize $N_\Delta$ as
$\lambda(t)$ grows. A  (projective) measurement detecting the presence
of a  dimer on  some bond crossing   $\Delta$ will thus  give $1$ with
probability $|\alpha|^2/L$.  The whole operation has  to be repeated a
large number of times  ($\sim L$) before $|\alpha|^2$  is known with a
reasonable accuracy  but   one  may  improve the efficiency    of  the
measurement by having a bond along $\Delta$ where the energy cost of a
dimer is   less  than on  other bonds   (in which  case the  dimer, if
present, will localize on this particular  bond when $\lambda$ becomes
large).  Of course the  reading procedure described above suffers from
the same limitations (time proportional to $L$ and low temperature) as
the unitary rotation.

\section{Conclusions}

We have shown that the QDM of Eq.~\ref{eq:h0} can  be simply solved in
the presence of two kinds of perturbations: an external potential that
couples to dimers crossing a line or  the inclusion of monomers.  This
provides a  simple example of system with  a $\Z$ fractionalized phase
where the  topological degeneracy is destroyed  by  tuning an external
parameter through   a   quantum phase  transition  (belonging   to the
classical Ising 2D universality class).

We also discussed some properties of this  toy model from the point of
view of an ideal topological  qubit, in which  case the exact solution
allows to follow  the  two lowest  eigenstates as  a function of  some
external  parameters.  These  two parameters can  be  used  to perform
unitary rotations of the qubit and provide an exactly solvable version
of  some  ideas  introduced previously.\cite{ioffe02} In  addition, we
pointed out that the phase transition could,  in principle, be used to
improve the robustness to decoherence because it could enable to use a
larger (although not infinite) system.   Concerning the measurement of
the qubit, we  also emphasize the  interesting properties of the phase
transition as  it  turns a non-local   property  ({\em parity} of  the
number  of dimers crossing a line)  into a local  property (dimer {\em
density}).  From this point of view, we  note that the method has some
resemblance with the flux trapping  experiment imagined by Senthil and
Fisher\cite{sf01} to detect visons in a $\Z$ fractionalized systems.

\section*{Acknowledgments}
We thank  Beno\^{\i}t  Dou{\c  c}ot  for stimulating  discussions  and
D.~Ivanov for useful  comments on the  manuscript. G.~M.  is  in part
supported   by the  Minist\`ere   de  la  Recherche  et  des Nouvelles
Technologies with an  ACI  grant and acknowledges the   hospitality of
IRRMA.

\appendix
\section{Ground-state energy of an ICTF with open boundary conditions}
\label{sec:eictfO}
We  apply fixed boundary conditions  to an  open  ICTF and compute the
energy difference between the case of ferromagnetic boundary conditions
(two   fixed up spins   at  the ends)  and antiferromagnetic  boundary
conditions (one up spin at one end and a  down spin at the other). This
result  has been   obtained    recently  by   Dou{\c c}ot   {\it    et
al.}\cite{dfii04} but   we  give here   for completeness  a   detailed
derivation of the result.

\subsection{Hamiltonian and free fermions}
The Hamiltonian is
\begin{equation}
        \mathcal{H}=
        -\sum_{n=1}^L\sigma^x_n
        -\mu\sum_{n=0}^{L} \sigma^z_n \sigma^z_{n+1}
\end{equation}
with two fixed spins at the ends of  the chain: $\sigma^z_{L+1}=1$ and
$\sigma^z_0=\pm1$  depending  on the  boundary  conditions.    Using a
standard Jordan-Wigner transformation to represent the Ising operators
with spinless fermions:
\begin{eqnarray}
  \sigma^x_n&=&2c^\dagger_n c_n -1
    \label{eq:sigmax}\\
  \sigma^y+i\sigma^z&=&
  2c^\dagger_n \exp{\left(i\pi\sum_{i=0}^{n-1}c^\dagger_i c_i\right)}
    \label{eq:sigma+}\\
  \sigma^z_n\sigma^z_{n+1}&=&
        (c^\dagger_n+c_n)(c_{n+1}-c^\dagger_{n+1})
  \label{eq:zzz}
\end{eqnarray}
$\mathcal{H}$ is quadratic in     the fermion operators and  can    be
diagonalized  by    a   Bogoliubov transformation.   To    find    the
quasi-particle  creation    operators $d^\dagger$  we    consider  the
following form   (Ansatz):\footnote{Eq.~\ref{eq:dff} will  insure that
$\sigma^z_0=c_0^\dagger+c_0$ do   not appear in  $d^\dagger_\omega$ so
that $d^\dagger_\omega$ and $d_\omega$ anticommute with $\sigma^z_0$.}
\begin{eqnarray}
    d^\dagger_\omega&=&f^\dagger_\omega-f^\dagger_{\omega^{-1}}
\label{eq:dff} \\
    f^\dagger_\omega&=&\sum_{n=0}^{L+1}
        \omega^n (c^\dagger_n+c_n)
    +ib(\omega)\sum_{n=0}^{L+1} \omega^n (c^\dagger_n-c_n)
\end{eqnarray}
where $\omega$ and $b(\omega)$ have to be determined.
One can check that
\begin{eqnarray}
    \left[\mathcal{H},d^\dagger_\omega\right]&=&E(\omega)d^\dagger_\omega \label{eq:Hd}\\
    \left[\mathcal{H},d_\omega\right]    &=&-E(\omega)d_\omega \nonumber \\
    {\rm with}\;\; E(\omega)&\geq&0 \nonumber
\end{eqnarray}
provided the following equations are satisfied:
\begin{eqnarray}
    E(\omega)&=&-2ib(\omega)(1+\lambda\omega^{-1}) \label{eq:E1} \\
    iE(\omega)b(\omega)&=&-2(1+\lambda\omega) \label{eq:E2} \\
    \omega^{L+1}-\omega^{-L-1}&=&-\lambda\left(\omega^{L+2}-\omega^{-L-2}\right)
    \label{eq:omega}
\end{eqnarray}
The  two   first equations  come from the   terms  $0\leq n\leq  L$ in
Eq.~\ref{eq:Hd}.    These   equations determine  the   energy  of  the
quasi-particles as a function of $\omega$:
\begin{equation}
E(\omega)=2\sqrt{\lambda^2+\lambda(\omega+\omega^{-1})+1}
\end{equation}
The third equation (\ref{eq:omega}) comes from the boundary
at $n=L+1$  and is a  constraint on the  available
$\omega$.

\subsection{Case $\lambda\geq \frac{L+1}{L+2}$}

For $\lambda\geq \frac{L+1}{L+2}$ the Eq.~\ref{eq:omega} has
$L+1$ distinct solutions of the form:
\begin{eqnarray}
    \omega&=&e^{ik}\;\;{\rm with}\;\;k\in\left]0,\pi\right] \nonumber \\
    \lambda&=&-\frac{\sin\left((L+1)k\right)}{\sin\left((L+2)k\right)}
    \label{eq:omegak}
\end{eqnarray}
These fermionic levels are those required to  describe the $L+1$ spins
$n=0\cdots L$. Because  we have chosen  $E$ to be always positive, the
absolute  ground-state  (whatever     $\sigma^z_0$) is   the    vacuum
$|0\rangle$ of  the $d^\dagger_\omega$.  The  lowest  excited state is
$d^\dagger_{k_0}|0\rangle$ where we   have added the fermion with  the
smallest   energy $\Delta$.  It corresponds  to  the solution $k_0$ of
Eq.~\ref{eq:omegak} which is the closest  to $\pi$. This solution  can
be calculated by an expansion in $1/L$ and we obtain
\begin{eqnarray}
\Delta&=&2(\lambda-1)+\frac{\lambda\pi^2}{(\lambda-1)L^2}+\mathcal{O}(L^{-3})
\label{eq:delta1}
\end{eqnarray}
So far  we have not specified  the value of $\sigma^z_0$ corresponding
to each level.  In the limit where $\lambda\gg1$ it  is clear that the
ground-state is in the sector $\sigma^z_0=1$.  Since one can show that
no level-crossing occur for $\lambda>0$   in such a finite chain,  the
fermion vacuum  $|0\rangle$  satisfies $\sigma^z_0|0\rangle=|0\rangle$
for all $\lambda>0$  and  correspond to   a state of  the system  with
ferromagnetic    boundaries.    On the    other  hand,  inserting  any
$d^\dagger_\omega$ fermion changes the sign of $\sigma^z_0$ (since all
the $d^\dagger_\omega$ anticommute with $\sigma^z_0=c_0^\dagger+c_0$).
The first excited state $d^\dagger_{k_0}|0\rangle$ is thus the
ground-state of the system  with antiferromagnetic boundary  conditions
($\sigma^z_0=-1$) and the  gap between  the  two sectors is  given  by
$\Delta$ (Eq.~\ref{eq:delta1}).

\subsection{Case $0<\lambda\leq \frac{L+1}{L+2}$}
In  the range $0>\lambda\geq  \frac{L+1}{L+2}$,  only $L$ solutions of
the  form  Eqs.~\ref{eq:omegak}    exist.  The   ``missing''  solution
$\omega_0$      has   the      lowest     energy      and is     real:
$\omega_0\in]-\infty,0]$. It corresponds  to  a bound-state (imaginary
wave-vector) for the  fermions. In  the  thermodynamic limit  one  has
$\omega_0=-\frac{1}{L}$ and finite-size corrections can be evaluated:
\begin{equation}
\omega_0=-\frac{1}{L}+\lambda^{2L+1}\left(1-\lambda^2\right)+\mathcal{O}(L\lambda^{4L})
\end{equation}
which gives an energy gap
\begin{equation}
\Delta=2\lambda^{L+1}\left(1-\lambda^2\right)+\mathcal{O}(L\lambda^{3L})
\label{eq:delta2}
\end{equation}
As before, this gap   is the  ground-state energy  difference  between
antiferromagnetic and ferromagnetic  boundary conditions  for the spin
chain.

\section{Ground-state energy of an ICTF  with periodic or antiperiodic boundary
conditions}
\label{sec:eictfC}
We compute the ground-state energy of a  closed ICTF with periodic and
anti-periodic boundary conditions.  The  latter  result has also  been
obtained recently by Dou{\c c}ot {\it et al.}\cite{dfii04} but we give
here a detailed  derivation  of the result.   We  also note that  this
calculation  has  some   similarities with  the   evaluation  of   the
ground-state  energy splitting in  the  triangular lattice QDM at  the
Rokhsar Kivelson point (using a Pfaffian technique).\cite{iif02}

\subsection{Periodic boundary conditions}
The Hamiltonian is :
\begin{equation}
        \mathcal{H}=
        -\sum_{n=0}^{L-1}\sigma^x_n
        -\mu\sum_{n=0}^{L-1} \sigma^z_n \sigma^z_{n+1}
\end{equation}
with   $\sigma^z_L=\sigma^z_0$.  The Ising  operators  are represented
with spinless fermions as in Eqs.~\ref{eq:sigmax}, \ref{eq:sigma+} and
\ref{eq:zzz}.  Due to the periodic boundary conditions we have also
\begin{eqnarray}
  \sigma^z_{L-1}\sigma^z_0&=&-(c^\dagger_{L-1}+c_{L-1})(c_0-c^\dagger_0)
  \nonumber \\
  &&\;\;\;\;\times\exp{\left(i\pi\sum_{n=0}^{L-1}c^\dagger_n c_n\right)}
  \label{eq:apbc}
\end{eqnarray}
It is simple to check that
\begin{equation}
\prod_{n=0}^{L-1}\sigma^x_n=\exp{\left(i\pi\sum_{n=0}^{L-1}c^\dagger_i c_i\right)}
        \label{eq:psnf}
\end{equation}
is a conserved quantity.  The spectrum can  thus be studied separately
in  the   sectors   $\prod_{n=0}^{L-1}\sigma^x_n=\pm1$.        However
$\mathcal{H}$ has off-diagonal  matrix element  in the  natural  Ising
basis which  are all $\leq0$.  This  insures that the ground-state has
only positive weight in this basis and it therefore belongs to sector
\begin{equation}
        \prod_{n=0}^{L-1}\sigma^x_n=1
        \label{eq:psz}
\end{equation}
In     the  following we thus  consider     fermions subjected to {\em
anti-periodic boundary conditions} (see Eqs.~\ref{eq:psnf} and
\ref{eq:psz} and the $-$ sign in the r.h.s of
Eq.~\ref{eq:apbc}).

After Fourier transform the Hamiltonian becomes:
\begin{eqnarray}
        \mathcal{H}_{\rm chain} &=&
        \sum_{k=\frac{2n+1}{L}\pi} \left[
        \left(i\mu\sin(k)c^\dagger_kc^\dagger_{-k}+{\rm H.c}\right)
        \right.\nonumber\\
        &&\left.
        -2c^\dagger_kc_k\left(\mu\cos(k)+1\right)
        +1
        \right]
\end{eqnarray}
which is diagonalized by a Bogoliubov transformation
\begin{eqnarray}
        \mathcal{H}_{\rm chain} =
        \sum_{k=\frac{2n+1}{L}\pi} \epsilon(k) \left[
        d^\dagger_kd_k-\frac{1}{2}
        \right] \\
                \epsilon(k)=2\sqrt{\mu^2+1+2\mu\cos(k)}
\end{eqnarray}
Using the explicit  form of the  transformation  and the anti-periodic
boundary conditions   on the fermions,  one can  show  that the vacuum
$|0\rangle$ of the $ d^\dagger_k$ fermions satisfies:
\begin{equation}
        \exp{\left(i\pi\sum_{n=0}^{L-1}c^\dagger_n c_n\right)}|0\rangle
        =+|0\rangle
\end{equation}
This is consistent with Eq.~\ref{eq:psz}  and the ground state is thus
$|0\rangle$.  Its energy is
\begin{equation}
        E_P=-\frac{1}{2}\sum_{k=\frac{2n+1}{L}\pi} \epsilon(k)
	\label{eq:EP}
\end{equation}
\subsection{Anti-periodic boundary conditions}

To    insure that    $\sigma^z_L=-\sigma^z_0$, the   fermions  are now
subjected   to     {\em    periodic}   boundary    conditions     (see
Eq.~\ref{eq:apbc}).   However, for $\mu>1$ it  is necessary to add one
$d^\dagger_k$ fermion to insure the correct parity under a global spin
flip.   Since the dispersion    relation $\epsilon(k)$ is  minimum  in
$\epsilon(k=0)=2|\mu-1|$ the ground-state for $\mu>1$ is
\begin{equation}
        |1\rangle=d^\dagger_0|0\rangle
\end{equation}
The  ground-state   energy of the  chain  with   anti-periodic boundary
conditions is thus
\begin{eqnarray}
        E_A     &=&-\frac{1}{2}\sum_{k=\frac{2n}{L}\pi} \epsilon(k)
                        +2(\mu-1)\;\; {\rm for}\;\;\mu>1
	\label{eq:EAmugt1}\\
                &=&-\frac{1}{2}\sum_{k=\frac{2n}{L}\pi} \epsilon(k)
                                 \;\;\;\; {\rm for}\;\;0\leq\mu\leq 1
\end{eqnarray}
\subsection{Energy difference}
\begin{figure}
  \begin{center}
    \includegraphics[width=6cm]{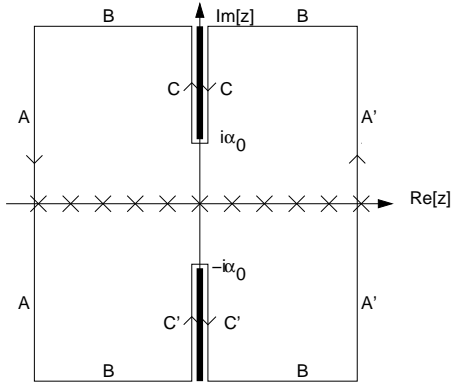}
    \caption[99]{
      Contour in the complex plane used to define the integral $I_C$.
      The crosses on the real axis indicate the poles of $1/\sin(Lz)$
      and the fat segments on the imaginary axis
      indicate the branch cuts of $f(z)$. The region B is sent
      to ${\rm Im}[z]=\pm\infty$.}\label{fig:contour}
  \end{center}
\end{figure}

From   the   calculation above   the   energy difference  between  the
ground-states of the two boundary conditions is (for $\mu\leq1$) :
\begin{eqnarray}
        E_A-E_P&=&\sqrt{2\mu} \sum_{n=0}^{L-1} \left[
        \sqrt{\cosh \alpha_0-\cos(k_{n+\frac{1}{2}})}\right.\nonumber\\
        &&\;\;\;\;\;\;\;\left.-\sqrt{\cosh \alpha_0-\cos(k_n)}\right]
\end{eqnarray}
where
\begin{eqnarray}
        k_n=\frac{2n\pi}{L}
\end{eqnarray}
and $\alpha_0$ is defined by
\begin{eqnarray}
        \cosh \alpha_0&=& \frac{\mu^2+1}{2\mu} \;\;\;(\geq 1)\\
        \alpha_0&=&-\ln(\mu)
\end{eqnarray}
The difference between the  two  sums  can  be  related to a   contour
integral $I_0$ in the complex plane:
\begin{equation}
                E_A-E_P=\sqrt{2\mu} \;L\; I_0
                \label{eq:EAEPI0}
\end{equation}
where
\begin{equation}
I_0=-\frac{1}{2i\pi}\oint_C \frac{f(z)}{\sin{(Lz)}}dz
\end{equation}
and
\begin{equation}
f(z)=\sqrt{\cosh \alpha_0-\cos(z)}
\end{equation}
The   contour   is   shown    Fig.~\ref{fig:contour}.  The    equality
Eq.~\ref{eq:EAEPI0} can  be  demonstrated by using  the  fact that the
poles     inside    the   contour   are   located     at   $z=k_n$ and
$z=k_{n+\frac{1}{2}}$   and have alternating  residues proportional to
$-f(k_n)/L$ and $+f(k_{n+\frac{1}{2}})/L$.
The   contour    can    be     decomposed   into   several    regions:
$I_0=I_A+I_{A'}+I_B+I_C+I_{C'}$  (Fig.~\ref{fig:contour}).  Using  the
odd parity of the integrand  under  $z\to-z$ one has $I_C=I_{C'}$  and
using     the   periodicity    under    $z\to      z+2\pi$ one   finds
$I_A+I_{A'}=0$. The    integrand  is exponentially  small   when ${\rm
Im}[z]\to\pm\infty$   so  the region  $B$   does  not contribute.   We
therefore  have $I_0=2I_C$. The integral over  the region $C$ is given
by the discontinuity of the integrand along the branch cut:
\begin{eqnarray}
  I_C&=&\frac{1}{2i\pi}\int_{\alpha_0}^\infty idr
  \left[
    \frac{f(ir+o^-)}{\sin(L(ir+o^-))}
  \right.\nonumber\\
  &&\;\;\;\left.
    -\frac{f(ir+o^+)}{\sin(L(ir+o^+))}
  \right] \nonumber \\
  &=&\frac{1}{\pi} \int_{\alpha_0}^\infty dr
  \frac{\sqrt{\cosh(r)-\cosh(\alpha_0)}}{\sinh(Lr)}
\end{eqnarray}
When the system size is large ($L\to\infty$) the behavior of $I_C$
is dominated by values of $r$ close to $\alpha_0$.
In this limit
\begin{eqnarray}
  I_C &\simeq&\frac{\sqrt{\sinh{\alpha_0}}}{\pi}
  \int_{\alpha_0}^\infty dr
  e^{-Lr}\sqrt{r-\alpha_0} \nonumber \\
  &\simeq&\frac{\sqrt{\sinh{\alpha_0}}}{2\sqrt\pi}
  e^{-L\alpha_0}L^{-\frac{3}{2}}
\end{eqnarray}
so that the energy difference is
\begin{eqnarray}
        E_A-E_P&=& 2 \sqrt{2\mu}\;L\;I_C \nonumber \\
        &\simeq& 2\sqrt{\frac{2\mu\sinh{\alpha_0}}{L\pi}}e^{-L\alpha_0}\\
        &\simeq&2\sqrt{\frac{1-\mu^2}{L\pi}}\mu^L
	\label{eq:delta3}
\end{eqnarray}

The  calculation of $E_A-E_P$ for  $\mu>1$ is almost  identical to the
$\mu<1$ case  described above.   The difference  between the  sums  of
$\epsilon(k)$ on 'even' and 'odd' momenta is  again expressed with the
integral $I_C$   but with  $\alpha_0=\ln(\mu)$.   Combining  this with
Eqs.~\ref{eq:EP} and \ref{eq:EAmugt1} gives:
\begin{equation}
	E_A-E_P=2(\mu-1)+2\sqrt{\frac{\mu^2-1}{L\pi}}\mu^{-L}
\end{equation}


\end{document}